\documentclass[proceedings]{rmaa}

\renewcommand{\P}[1]{%
\ifnum#1=1\hbox{OW~168--326E}\fi
\ifnum#1=2\hbox{OW~167--317}\fi
\ifnum#1=3\hbox{OW~163--317}\fi
\ifnum#1=5\hbox{OW~158--323}\fi
\ifnum#1=0\hbox{OW~171--334}\fi}

\title{Constraints on Type Ia Supernovae from Near Infrared Spectra}
\author{G. H. Marion, P. H\"oflich, $\&$ J. C. Wheeler 
  \affil{Department of Astronomy, University of Texas at Austin} }

\fulladdresses{
\item G. H. Marion, P. H\"oflich, $\&$ J. C. Wheeler,
  Department of Astronomy, University of Texas at Austin,
  Austin, TX, 78712 (hman@astro.as.utexas.edu).}

\shortauthor{Marion, H\"oflich $\&$ Wheeler}
\shorttitle{SNe Ia Constraints from NIR Spectra}


\abstract{  We describe a research program to improve the understanding of Type Ia Supernovae (SNe Ia) by modeling and observing near infrared (NIR) spectra of these events.  The NIR between 0.9$\mu$ and 2.5$\mu$ is optimal for examining certain products of the SNe Ia explosion that may be blended or obscured in other spectral regions.  NIR analysis will enable us to place important constraints on the physical properties of SNe Ia progenitors and their explosion dynamics. These are critical steps toward understanding the physics of Type Ia Supernovae.  

We have identified features in NIR spectra of SNe Ia that discriminate between Population I and Population II progenitors.  These features can significantly restrict the evolutionary history of SNe Ia.  We also examine certain products of the nuclear burning that enable us to place constraints on the propagation of nuclear burning during the explosion, and on the behavior of the burning front during the event.  We will be able to differentiate between the several explosion models for SNe Ia.
}

\listofauthors{G. H. Marion, P. H\"oflich, $\&$ J. C. Wheeler}
\indexauthor{Marion, G. H.}
\indexauthor{H\"oflich, P.}
\indexauthor{Wheeler, J. C.}

\begin{document}
 
\maketitle

\section{Introduction}
\label{sec:intro}
The tremendous brightness and apparent homogeneity of SNe Ia events make them excellent probes of the universe at great distances. Type Ia supernova observations have provided high quality data for estimates of the Hubble constant (M\"uller \& H\"oflich 1994, Nugent et al. 1996), analysis of the matter density of the universe (H\"oflich, Wheeler, \& Thielemann 1997 and references therein), and possibly demonstrated a positive value for the Cosmological Constant (Riess et al. 1998, Perlmutter et al. 1999). The conclusion that the expansion rate of the universe is increasing was declared the most significant scientific discovery of 1998 by Science Magazine.  Those results however, are based entirely on empirical analysis of light curve observations without any physical foundation for the behavior.  New methods of testing the cosmological implications from the observational data of SNe Ia are required.  In addition to studying SNe Ia physics and evolution, our research is an important step toward understanding the systematics and calibration of the high-z supernova searches.

The observed characteristics of SNe Ia strongly suggest that these events are the thermonuclear explosions of Carbon/Oxygen White Dwarf (WD) stars (H\"oflich \& Khokhlov 1996 and references therein).  The canonical model for SNe Ia consists of a C/O WD accreting matter from a binary companion.  As the WD approaches the Chandrasekar limiting mass, energy from surface burning of accreted material and compressional heating ignites a thermonuclear runaway.  This explosion mechanism distinguishes SNe Ia from all other supernovae, which are thought to be core collapse events.  Although this general scenario is accepted, the details of the explosion and the progenitor remain under discussion.

\section{Models} 
\label{sec:Models}
We are working with  hydrodynamic simulations of the supernova explosion and a code for nucleosynthesis and radiation transport that have been very successful in making predictions and analyzing the optical behavior of SNe Ia (H\"oflich \& Khokhlov 1996).  This code runs locally on a 40 processor parallel Beowulf system.

The event dynamics follow a Deflagration to Detonation Transition (DDT) explosion model.  A prompt detonation mechanism for SNe Ia would result in nearly all of the stellar material being burned directly to iron and heavier products.  That is inconsistent with observations of intermediate mass elements in SNe Ia spectra near maximum light.  An expansion phase is required prior to the explosion in order to allow burning at lower densities and temperatures than would be possible in a prompt detonation.  Our models include an initial subsonic deflagration wave that becomes a detonation shock at a critical density.  After transition, the burning front proceeds by compressional heating at velocities near to the speed of sound.  A deflagration front that propagates at very nearly sonic speed may produce similar results.  Soon after the shock reaches the surface of the WD, the explosion products continue in homologous expansion.

As the layers of the supernova ejecta expand, the photosphere recedes in mass through these layers.  Once the photosphere has moved entirely within the envelope containing a particular element, we continue to observe the expansion velocity of the inner edge of that layer until the density is reduced to the point that the element no longer detectable.  In the models, we calculate the radial velocities of the explosion products and use them to construct the synthetic spectra.  In observation, radial velocities of the elements are determined from the Doppler shifts of their spectral lines.  

\section{NIR Spectra} 
\label{sec:spectra}

The model spectra we produce reveal products of the explosive nucleosynthesis and a few trace elements from the progenitor.  Among the metal lines that appear in NIR spectra is a MgII line near 1.05$\mu$.  Magnesium is a product of explosive carbon burning, but not oxygen burning for which the higher temperatures generate products further along the alpha chain.  That makes magnesium an excellent diagnostic tool for defining the region of transition between carbon and oxygen burning during the explosion.  This transition is a key aspect of explosion models of SNe Ia that require both a deflagration (subsonic) burning front to produce the observed intermediate mass elements, and a detonation (supersonic) burning front to produce sufficient energetics.  We will use this MgII line to probe the explosion dynamics as well as the propagation of the deflagration burning front through the WD. 

Another NIR feature is the FeII line at 1.00$\mu$.  When it is detected near maximum light, this line is an indicator of the metallicity of the progenitor WD.  In SNe Ia, electron capture is significant only within the inner layers.  The outer layers retain the initial electron to nucleon ratio ($Y_e$) with which the progenitor was created.  For a WD with zero metallicity, $Y_e=0.5$ and that value is reduced as metallicity increases.  Before maximum light in SNe Ia, the only iron available to make the 1.00$\mu$ FeII line will be $^{54}Fe$ generated by incomplete $O$ burning with $Y_e<.495$.  Soon after maximum light however, this line will be dominated by $^{56}Fe$ which is the decay product from $^{56}Ni$ through $^{56}Co$.  We predict that this FeII line will be prominent before maximum light in SNe Ia spectra from progenitors that have solar metallicity or greater, but will be undetectable in spectra from low metallicity progenitors.   

The ability to determine metallicity in the progenitor will create a new tool to explore the characteristics of SNe Ia as a function of redshift and of the morphology of the host galaxy. This result is critical to the use of SNe Ia for cosmology because it permits study of the systematic effects from the comparison of distant supernovae to nearby events.  Alternatively, a low electron to nucleon ratio may be achieved by a large number of Helium shell flashes in the progenitor evolution.  We also study other features in NIR spectra, including the CaII line at 1.2$\mu$, and the SiII line at 1.7$\mu$.

Comparing two of our models to data from SN1994D (Meikle 1996) demonstrates the sensitivity of our technique (See Figure 1).  The models are identical except for a 10$\%$ difference in the density at which the nuclear burning front makes the transition from a subsonic deflagration to supersonic detonation.  The difference in transition density, produces a change the radial velocity at the inside edge of the magnesium envelope of only 1000 km/s, but we are able to easily observe the change in our model spectra.  The range of minimum velocities for MgII in reasonable SNe Ia models is about 5000 km/s.  Our ability to distinguish a 1000 km/s difference makes this a very accurate diagnostic tool.

\begin{figure}[ht]
  \begin{center}
    \leavevmode
    \includegraphics[width=5.0in]{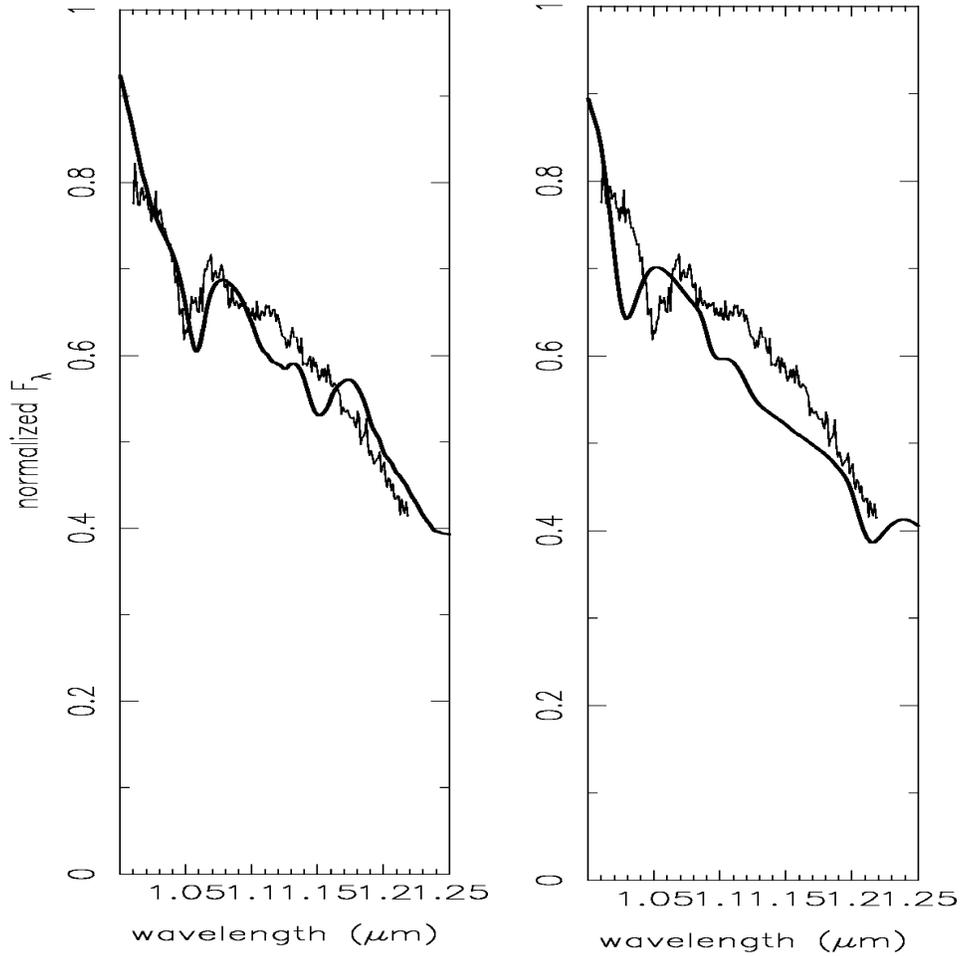}
    \caption{
The figures show synthetic spectra produced by our models (smooth line) superimposed on data from SN1994D (Meikle 1996).  The vertical axis is normalized flux, and the horizontal axis is wavelength in microns (divisions of .01$\mu$ are labeled at: 1.05, 1.1, 1.15, 1.2 \& 1.25).  The two models are identical except for a 10$\%$ difference in the density at which transition occurs from deflagration burning (subsonic) to detonation burning (supersonic).  This change in transition density produces a change in the minimum radial velocity for magnesium of about 1000 km/s.  The MgII line can be seen in the data as a P-Cygni profile with a blue-shifted absorption minimum at about 1.05$\mu$.  The difference in the accuracy of the fit between the models and the data is apparent.  Our models can easily identify a 1000 km/s difference in radial velocity.}
   \label{fig:spectra}
  \end{center}
\end{figure}

\section{Observation}
\label{sec:observation}
An important goal of this program is to increase the available data on SNe Ia in the NIR.  We currently have four NIR spectra from SNe Ia: 1986Y, 1994D, 1998bu, $\&$ 1999by.  Additional observations are required for statistical completeness and to cover a range of peak absolute magnitudes.  The metal lines we want to observe will be available from a few days after the explosion until about 1 week after maximum light.  The window of opportunity will depend on discovery time, but typically we will have approximately two weeks to make the observations.  An analysis of supernova discoveries since January 1, 1999 (CBAT 2000), suggests that for an instrument and telescope combination with a limiting magnitude of 16.5, there is a better than even chance that on any given night there will be a SNe Ia event within 15 days of discovery available for observation.  Current SNe Ia discovery programs are optimized for slightly dimmer events and for a limiting magnitude of 17.5 it is 95$\%$ likely that at least one SNe Ia target will be available and often there will be more than one.

\section{Conclusion}
\label{sec:conclusion}
Our program will constrain SNe Ia physics by comparing model spectra between 0.9$\mu$ and 2.5$\mu$ to data from SNe Ia observations.  Investigation in the NIR allows us to monitor certain elements that are indicators of SNe Ia behavior but are difficult to observe in other spectral regions.  The models will cover a range of progenitor compositions and explosion dynamics and we have shown that they are responsive to small changes in parameter space.  Constraints we provide on progenitors will improve the understanding of SNe Ia evolution.  Additional constraints on explosion dynamics will have implications for the cosmological use of SNe Ia data as well as contributing to the understanding of the physics of Type Ia Supernovae.

\acknowledgements 
\noindent
{\it Acknowledgments:} We would like to thank R. Fesen, C. Gerardy, P. Meikle, E. Bowers \&  J. Spyromillo  for providing IR-data.  This research was supported in part by  NASA Grant LSTA-98-022, NSF Grant 9818960, and a grant from the Texas Advanced Research Program.

The calculations for the explosion and light curves were done on a cluster of
workstations financed by the John W. Cox-Fund  of the Department of Astronomy at the University of Texas, and processors donated by AMD.

\end{document}